\documentclass[dvips]{acta}
\usepackage{lscape,epsfig}
\usepackage{amssymb}
\usepackage{amsmath}
\usepackage{rotating}
\usepackage{multirow}

\SetPages{0}{0}

\SetVol{66}{2016}

\begin{document}

\newcommand{\TabCapp}[2]{\begin{center}\parbox[t]{#1}{\centerline{
  \small {\spaceskip 2pt plus 1pt minus 1pt T a b l e}
  \refstepcounter{table}\thetable}
  \vskip2mm
  \centerline{\footnotesize #2}}
  \vskip3mm
\end{center}}

\newcommand{\TTabCap}[3]{\begin{center}\parbox[t]{#1}{\centerline{
  \small {\spaceskip 2pt plus 1pt minus 1pt T a b l e}
  \refstepcounter{table}\thetable}
  \vskip2mm
  \centerline{\footnotesize #2}
  \centerline{\footnotesize #3}}
  \vskip1mm
\end{center}}

\newcommand{\MakeTableee}[4]{\begin{table}[htb]\TabCapp{#2}{#3}
  \begin{center} \TableFont \begin{tabular}{#1} #4
  \end{tabular}\end{center}\end{table}}

\begin{Titlepage}
\Title{OGLE-LMC-ECL-09937: The Most Massive Algol-Type Binary System With A Mass Measurement Accurate to $2\%$}
\Author{D.~M.~~S~k~o~w~r~o~n$^1$,
M.~~K~o~u~r~n~i~o~t~i~s$^{2,3}$,
J.~L.~~P~r~i~e~t~o$^{4,5}$,
N.~~C~a~s~t~r~o$^{6}$,
A.~Z.~~B~o~n~a~n~o~s$^{3}$,
D.~K.~~P~i~e~\'n~k~o~w~s~k~i$^1$}
{$^1$ Warsaw University Observatory, Al.~Ujazdowskie~4, 00-478~Warszawa, Poland\\
e-mail: dszczyg@astrouw.edu.pl\\
$^2$ Astronomick\'{y} \'{u}stav AV\v{C}R, Ond\v{r}ejov, 25165, Czech Republic\\
$^3$ IAASARS, National Observatory of Athens, Penteli 15236, Greece \\
$^4$ N\'ucleo de Astronom\'ia de la Facultad de Ingenier\'ia, Universidad Diego Portales,\\ Av. Ej\'ercito 441, Santiago, Chile\\
$^5$ Millennium Institute of Astrophysics, Santiago, Chile\\
$^6$ Department of Astronomy, University of Michigan, 1805 S.University, Ann Arbor,\\ MI 48109, USA
}
\Received{December 11, 2017}
\end{Titlepage}

\Abstract{
This paper presents a detailed analysis of the light and radial velocity curves 
of the semi-detached eclipsing binary system OGLE-LMC-ECL-09937. The system is 
composed of a hot, massive and luminous primary star of a late-O spectral type, 
and a more evolved, but less massive and luminous secondary, implying an 
Algol-type system that underwent a mass transfer episode. We derive masses of 
$21.04\pm0.34$ M$_{\odot}$ and $7.61\pm0.09$ M$_{\odot}$ and radii of $9.93\pm0.06$
R$_{\odot}$ and $9.18\pm0.04$ R$_{\odot}$, for the primary and the secondary 
component, respectively, which make it the most massive known Algol-type system 
with masses and radii of the components measured with $<2\%$ accuracy. 
Consequently, the parameters of OGLE-LMC-ECL-09937 provide an important
contribution to the sparsely populated high-mass end of the stellar mass
distribution, and an interesting object for stellar evolution studies, being a
possible progenitor of a binary system composed of two neutron stars.} 
{binaries: eclipsing -- Stars: early-type -- Stars: fundamental
parameters -- Stars: massive}

\section{Introduction}

Mass is the most important property of a star. The initial mass and
metallicity to a large extent determine the whole life and fate of a
star. Moreover, the evolution of the star in a binary system is
significantly influenced by its companion (Sana \etal 2012). Most
stars with well-determined masses are Galactic stars with roughly
solar metallicity. Obtaining accurate parameters for stars in
environments with different metallicities is crucial for testing
stellar evolutionary theories and for measuring distances to nearby
galaxies (\eg Bonanos \etal 2006). Moreover, investigating stars in
the lowest metallicity environments is a step toward exploring the
early Universe and puts us closer to the Population III regime of
metal-free stars (even if we lack true zero metallicity stars). The
stellar mass distribution, particularly the initial mass function
(IMF), is fundamental for constructing accurate models of galaxies and
tracing their evolution.

The high-mass end of the stellar mass distribution is very poorly
constrained, which is a serious problem for verifying theoretical
models of stellar evolution. At the same time, the evolution of
massive stars is violent and their deaths lead to many interesting
phenomena. For example, massive binary stars are thought to be the
progenitors of Gamma Ray Bursts, especially in the case of
Population~III stars (\eg Toma \etal 2016). Furthermore, procuring
accurate fundamental measurements of the masses of high-mass stars is
essential for determining the slope of the IMF.

The review by Torres \etal (2010) lists all detached binary systems of
any mass in which the mass and radius of both components are measured
with $\leq3\%$ accuracy. This is the precision necessary to test
models of stellar evolution, and provide constraints strong enough
that models with incorrect underlying physics can be rejected. The
list contains 190 individual stars, of which only two are
extragalactic. The most massive star on the list is a component of
V3903~Sgr which has a mass of only 27.27~\MS, and there are only three
stars with masses greater than 20~\MS\ and 17 with masses above
10~\MS. DEBCat (Southworth 2015) is a living catalog that provides
physical properties of stars in detached eclipsing binary systems that
have been determined with $\leq2\%$ accuracy. As of today, the list
contains 390~stars (195~systems) and the record still belongs to
V3903~Sgr. The number of stars with masses greater than 20~\MS\ is
four, while with masses greater than 10~\MS\ is 22. For stars with
$M>30$~\MS\ the situation is even worse, there are very few
measurements accurate to less than 10\% (\eg Bonanos 2009).

Motivated by the deficiency of massive stars with accurate mass
measurements in low metallicity environments, we undertook a
spectroscopic survey of the most luminous stars in the Large
Magellanic Cloud (LMC), with an aim to measure their masses with at
least 2\% accuracy, necessary for precise astrophysical studies. The
targets are the eight most luminous binary systems in the LMC
identified by Szczygie\l~\etal (2010). In this paper, we present the
analysis of the first candidate massive binary system from our sample,
namely BI~98~=~OGLE-LMC-ECL-09937.

This paper is organized as follows. Section~2 gives an overview of the
observational data and its preparation. In Section~3 we present the
process of estimating physical parameters of the system, and discuss
the results in Section~4. We summarize the paper in Section~5.

\section{Observations and Data Preparation}

OGLE-LMC-ECL-09937 was identified by Szczygie\l~\etal (2010) in the All
Sky Automated Survey (ASAS, Pojma\'nski 1997) data as an eclipsing
binary star ASAS 051230-6727.3. It is located in the LMC at
$\alpha=78\zdot\arcd1266632$ and $\delta=-67\zdot\arcd4539948$, has a
period of 3.765184~d, magnitude $V=14.03$~mag and amplitude of
0.62~mag. It has been assigned a B1:V spectral type by Massey \etal
(1995) and was included in the infrared study of massive stars by
Bonanos \etal (2009).

\subsection{Photometric data}

The original ASAS (Pojma\'nski 2002) light curve of ASAS051230-6727.3
that led to its identification as an eclipsing binary (Szczygie\l~\etal
2010), is of very low quality. However, the catalog of eclipsing
binaries from the OGLE survey (Graczyk \etal 2011) contains the high
quality {\it I}- and {\it V}-band light curves of the object,
extracted from the third phase of the OGLE project (Udalski \etal
2008). There are 476 and 40 points in the {\it I}- and {\it V}-band
light curves of OGLE-LMC-ECL-09937, respectively. The data were
reduced and calibrated with the standard OGLE image subtraction
pipeline (Udalski 2003). Thanks to the high quality of the OGLE data
we were able to refine the orbital period using the {\sc Tatry} code
based on the multi-harmonic periodogram of Schwarzenberg-Czerny (1996)
and the new value is $P=3.765229$~d. The corrected value of the time
of primary eclipse is ${\rm HJD}_0=2453637.487389$~d.

\subsection{Spectroscopic data}

Spectroscopic observations were carried out with the Du~Pont 2.5~m and
the Magellan Clay 6.5~m telescopes at the Las Campanas Observatory in
Chile. In the Du~Pont we used the Echelle spectrograph with a
$1\arcs\times4\arcs$ slit and $2\times2$ binning, yielding a
resolution of $R=30\,000$, which corresponds to 10~km/s, sufficient
for velocity separations of $\approx100$~km/s. Separate exposures were
collected, in order to facilitate the removal of cosmic rays and
artifacts. For the wavelength calibration, ThAr lamp spectra of 20~s
were taken before and after each science exposure. Milky Flats and
Bias frames were collected in the afternoon preceding each
night. Typical signal to noise is about 60 for the combined
spectra. In the Magellan Clay we used the Magellan Inamori Kyocera
Echelle (MIKE; Bernstein \etal 2003) spectrograph with a 1\arcs slit
and $1\times1$ binning, yielding a resolution of $R=30\,000$. For
calibration we obtained bias frames, quartz and milky flats, and ThAr
lamps for each science exposure. Table~1 provides the log of
spectroscopic observations.

\def\arraystretch{1.3}
\MakeTable{|c|c|c|r|c|c|c|}{12.5cm}{Spectroscopic observations of LMC3.}
{\hline 
 DATE & HJD & PHASE & EXPTIME & AIRMASS & S/N & TELESCOPE \\
\hline
2009-12-24 & 2455189.664040 & 0.24 &   $1200$ s      & 1.28 & 73 & Clay/MIKE\\
2010-10-08 & 2455477.770863 & 0.75 & $3\times1800$ s & 1.41 & 35 & Du~Pont/Echelle \\
2010-10-10 & 2455479.823712 & 0.30 & $3\times1800$ s & 1.31 & 35 & Du~Pont/Echelle \\
2010-11-09 & 2455509.705694 & 0.24 & $3\times1800$ s & 1.36 & 45 & Du~Pont/Echelle \\
2011-09-02 & 2455806.872237 & 0.16 &    $700$ s      & 1.38 & 58 & Clay/MIKE\\
2014-01-08 & 2456665.570596 & 0.22 &   $2400$ s      & 1.31 & 61 & Clay/MIKE\\ 
2017-01-10 & 2457763.565667 & 0.83 & $3\times1800$ s & 1.31 & 55 & Du~Pont/Echelle \\
2017-01-11 & 2457764.687263 & 0.13 & $3\times1800$ s & 1.34 & 50 & Du~Pont/Echelle \\
2017-01-12 & 2457765.592431 & 0.37 & $3\times1800$ s & 1.28 & 38 & Du~Pont/Echelle \\
\hline
\multicolumn{7}{p{9cm}}{HJD is calculated for the mid-exposure of the total
of all three frames}
}
\def\arraystretch{1.0}

The Du Pont Echelle data were reduced using standard IRAF (Tody 1986,
Tody 1993) packages.  First, a master Bias frame was constructed from
single Bias frames and subtracted from each science observation. Then
a normalized Milky Flat was created from a median of single images,
and corrected for bad pixels.  Finally, science spectra were divided
by the master Milky Flat. An initial cosmic ray rejection on single
images was performed using the DCR program (Pych 2004), which works
well with spectroscopic data.  The Clay MIKE data were reduced using
the Carnegie pipeline\footnote{\it
  http://code.obs.carnegiescience.edu/mike}, which uses similar
reduction steps as described for the Du Pont Echelle reduction.

The FITS images were further processed with the {\sf noao.imred.echelle} package
from IRAF. Both science and ThAr spectra were extracted with the task {\sf
apall}. The line identification in ThAr spectra was done with the {\sf
ecidentify} and {\sf ecreidentify} tasks and the wavelength calibration of
science spectra was achieved with {\sf refspectra} and {\sf discpcor} tasks. The
spectra were normalized with the {\sf continuum} task and merged with {\sf
scombine}. Finally, a barycentric correction was applied with the use of {\sf
rvsao.bcvcorr}.

\section{Binary Modeling}

\subsection{Spectroscopy and radial velocities}

The spectrum of OGLE-LMC-ECL-09937 is characterized by the presence of
neutral helium and hydrogen lines in absorption. Both components are
distinguishable, and most significant spectral lines are pictured in
Figs.~1 and~2 at different orbital phases. We classify the stronger,
primary component following the criteria and the optical atlas for OB
stars by Walborn and Fitzpatrick (1990). The weak, questionable HeII
4200~\AA\ implies that the~ star~~ is of late-O/early-B~~ spectral type.
\begin{landscape}
\begin{figure}[p]
\centerline{\includegraphics[width=20cm, bb=-250 100 780 675, clip=]{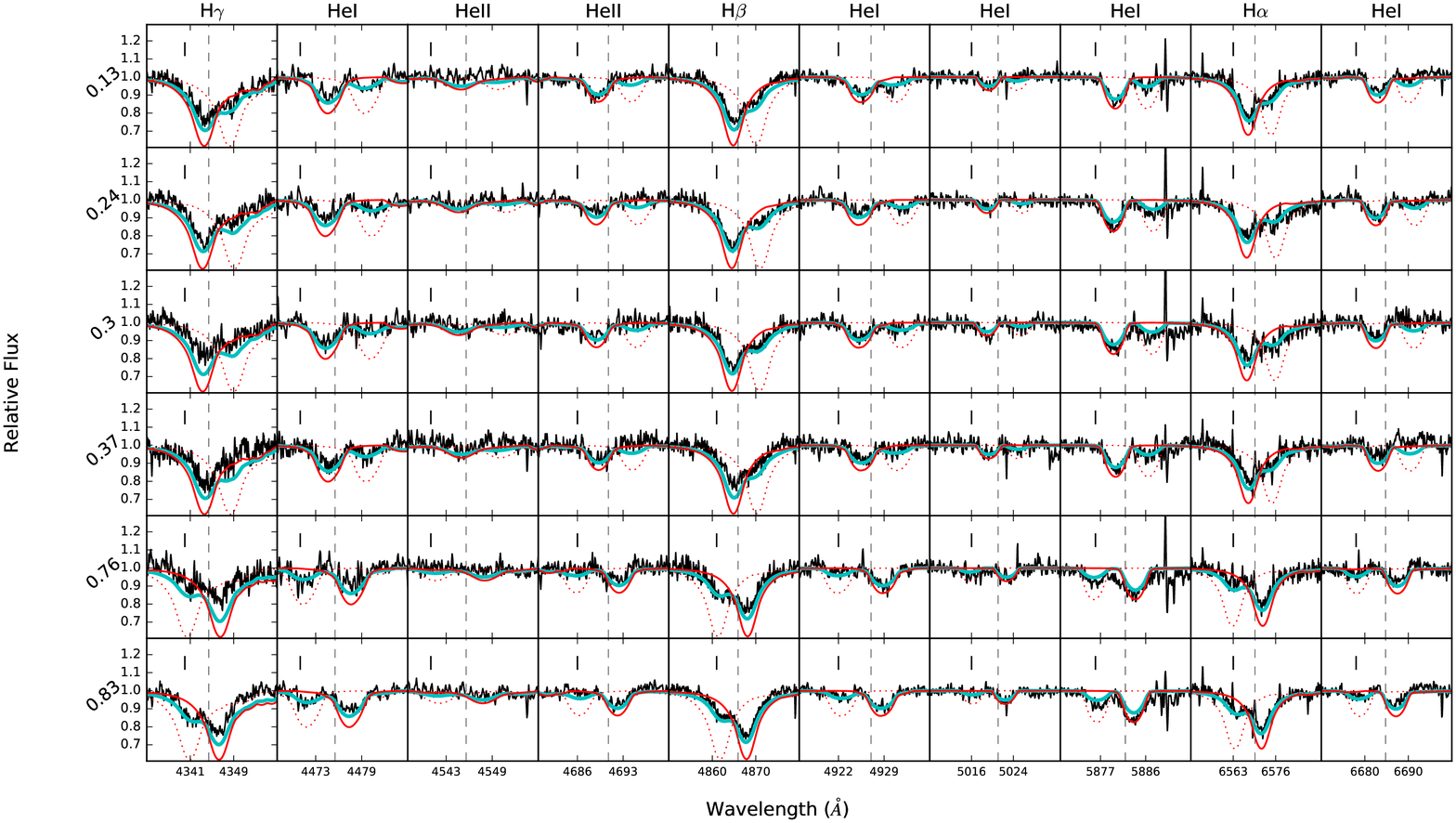}}
\FigCap{Selected H and He features in Du~Pont spectra of
  OGLE-LMC-ECL-09937 shown at different orbital phases, marked on the
  left. The red solid and dotted lines correspond to the primary and
  secondary component, respectively, shifted at their measured
  velocities given in Table~2. The combined fit to the observations
  (black line) is shown with a cyan solid line. The vertical solid
  line marks the rest wavelength of the spectral line, while the
  vertical dashed line, the location of the systemic velocity.}
\end{figure}
\end{landscape}
\noindent
The low ratio HeII 4541/HeI 4471 indicates a spectral type later than
O7, whereas the presence of HeII 4686 in absorption is typical for
high-gravity stars earlier than B0.5. Based on these diagnostics, the
primary component can be assigned a late-O main-sequence spectral
type. Absorption of HeII lines is not evident for the companion thus
favoring a B-type classification, although its small contribution to
the total flux prevents identification of intrinsically weak lines.

\begin{figure}[htb]
\centerline{\includegraphics[width=13cm]{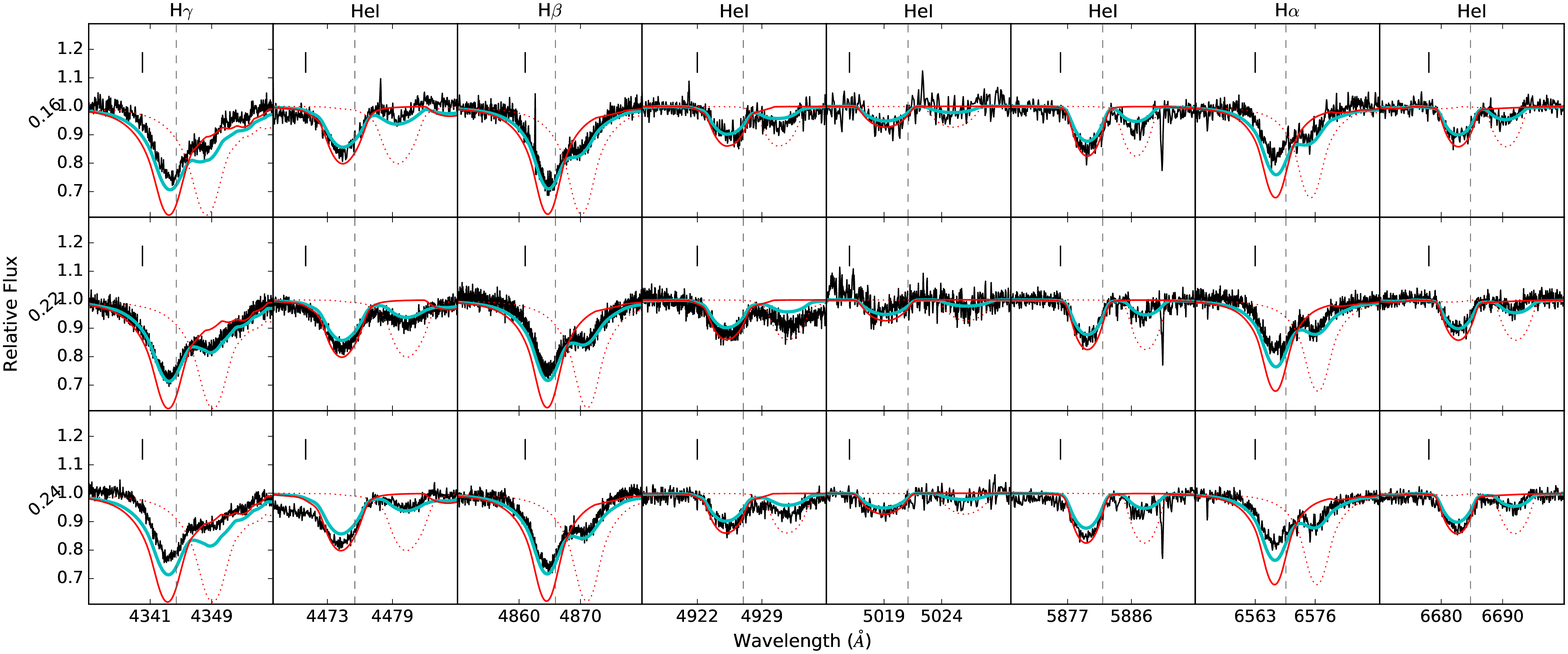}}
\vspace{-0.1cm}
\FigCap{Same as Fig.~1, but for the MIKE spectra. The four first
  features are from the BLUE arm and the four last from the RED arm
  of MIKE. }
\end{figure}

We proceeded to measure radial velocities (RVs) and estimate
spectroscopic temperatures in two steps. First, we pursued a
preliminary fit to the kinematics of the system and inferred stellar
properties that would physically constrain the final selection of
spectroscopic models. In a second step, we derived spectroscopic
temperatures and fine-tuned the RVs throughout the phases.

We employed a set of plane-parallel, NLTE, TLUSTY high-gravity models
(Hubeny and Lanz 1995) at $Z=0.5$~Z$_{\odot}$, with T$_{\rm
  eff}=30\,000$~K consistent with the late-O/early-B spectral type of
the system. To fit the observed spectra, we run a {\sf least-square
minimization} algorithm under a Levenberg-Marquardt scheme, setting
free the RVs of the components and the flux ratio. The algorithm was
run over spectral features visible for both components, namely H lines
(4340~\AA, 4861~\AA, 6563~\AA) and HeI lines (4471~\AA, 4922~\AA,
5016~\AA, 5876~\AA, 6678~\AA), which were selected due to their
strength and S/N ratio. The resulting RV curves along with the OGLE
light curves were then processed with PHOEBE v1.0 (Pr\v{s}a and
Zwitter 2005), which wraps the renowned Wilson-Devinney code for
modeling the physics of eclipsing binaries (Wilson ans Devinney 1971).

The system was initially modeled in the ``detached'' mode. The fit
solution however, converged to the semi-detached configuration with
the secondary component filling the Roche lobe, hence we switched to
that configuration. A by-eye reasonable fit of the light curves was
achieved from the manual adjustment of the surface potential of the
primary component $\Omega_1$, effective temperature of the secondary
$T_{\rm eff_2}$, and inclination $i$. The mass ratio $q$, semi-major axis
$a$, and systemic velocity $\gamma$ were adjusted to optimally fit the
RV curves. Bolometric and passband coefficients for the logarithmic
limb-darkening law were taken from Van Hamme (1993). After choosing
the initial values, all six parameters were set free, to allow for the
built-in fit optimization of PHOEBE based on the method of
differential corrections. The period and HJD$_0$ were fixed to the
values determined from the light curve analysis.

The best-fit PHOEBE model yielded surface gravity values of 3.75 and
3.40 for the hot and cool component, respectively. The resulting {\it
  V}-band light ratio of 2.3 was used to fix the spectroscopic flux
ratio. Having stellar models constrained by these values, we repeated
the RV velocity optimization using the Du~Pont observations of
2017-01-10 with the highest S/N, now setting the temperature of the
models free over a grid of TLUSTY models with $T_{\rm
  eff}=19\,000{-}40\,000$ K. Additionally, the models were broadened
over a wide range of rotational velocity values from 50~km/s to
200~km/s, with a step of 20~km/s.

We found that minimization of the residuals of the selected spectral
features was achieved with a primary component temperature $T_{\rm
  eff_1}=32\,500$~K. The temperature of the secondary was found to be
similar, but poorly constrained. Broadening of the primary was
optimized at ${\rm v}\sin i=150$~km/s. Because of the 2500~K step of the
TLUSTY model grid, we repeated the same process using models generated
with FASTWIND stellar atmosphere code (Santolaya-Rey \etal 1997, Puls
\etal 2005), which enables non-LTE calculations assuming a spherical
symmetry geometry. We generated a grid of FASTWIND templates with a
step of 1000~K at the inferred broadening with TLUSTY and using the
gravity values from the preliminary study with PHOEBE. The best fit
was achieved again with a primary temperature of $32\,000\pm1600$~K
and a secondary of temperature $34\,000\pm4000$~K. Interestingly, by
running the fit process for every spectrum, we derived a consistent
$T_{\rm eff_1}$ throughout the orbital phases, whereas $T_{\rm eff_2}$
was largely deviating within errors from 21\,000~K to even 40\,000~K
due to its low strength in the composite spectrum. We decided to infer
$T_{\rm eff_2}$ from the PHOEBE model and measure RVs using two
templates of 32\,000~K. The resulting values are listed in Table~2
with the best-fit model displayed over the observations in Figs.~1
and~2.

\def\arraystretch{1.3}
\MakeTableee{|c|c|c|c|c|}{12.5cm}{Radial velocity measurements for OGLE-LMC-ECL-09937}
{\hline
HJD & RV$_1$ (km/s) & RV$_2$ (km/s) & PHASE & TELESCOPE \\
\hline
\multirow{2}{*}{2455189.664040} & 194 (4) & 599  (8) & \multirow{2}{*}{0.24} & Clay/MIKE BLUE\\
                                & 198 (4) & 618  (8) &                       & Clay/MIKE RED\\
\hline
2455477.770863 & 419 (4) &  -2 (10) & 0.76 & Du~Pont/Echelle\\
\hline
2455479.823712 & 196 (4) & 589 (10) & 0.30 & Du~Pont/Echelle\\
\hline
2455509.705694 & 195 (4) & 599  (8) & 0.24 & Du~Pont/Echelle\\
\hline
\multirow{2}{*}{2455806.872237} & 206 (4) & 546  (8) & \multirow{2}{*}{0.16} & Clay/MIKE BLUE\\
                                & 207 (6) & 557 (12) &                       & Clay/MIKE RED\\
\hline
\multirow{2}{*}{2456665.570596} & 200 (4) & 603  (4) & \multirow{2}{*}{0.22} & Clay/MIKE BLUE\\
                                & 207 (4) & 628  (6) &                       & Clay/MIKE RED\\
\hline
2457763.565667 & 403 (4) &  29  (8) & 0.83 & Du~Pont/Echelle\\
\hline
2457764.687263 & 215 (4) & 537  (8) & 0.13 & Du~Pont/Echelle\\
\hline
2457765.592431 & 218 (4) & 550  (6) & 0.37 & Du~Pont/Echelle\\
\hline
\noalign{\vskip5pt}
\multicolumn{5}{p{10cm}}{Uncertainties ($2\sigma$) are given in parentheses.}
}

\subsection{Fundamental parameters of OGLE-LMC-ECL-09937}

Having measured the radial velocities, we used them together with the
high quality OGLE-III {\it V}- and {\it I}-band light curves to
calculate the final fit parameters with PHOEBE. Since the
uncertainties of the parameters provided by the {\sc gui} interface of
PHOEBE are underestimated, we used the {\sf scripter} to estimate their
realistic values. First, we generated a 1000 bootstrap samples of both
the light and radial velocity curves. The light curves were created by
$k$-multicombination of the original light curve consisting of $k$
points (where $k$ is 476 and 40 for the {\it I}- and {\it V}-band,
respectively). Due to the small number of points, the RV curves were
created by randomly selecting each point within its $2\sigma$
uncertainty around the measured value (listed in Table~2), assuming a
Gaussian distribution. The temperature of the primary was also
randomized over its spectroscopic value $32\,000\pm1600$~K. We then
ran PHOEBE on each of the 1000 bootstrap samples and derived 1000 sets
of parameters of the system. The mean values of the parameters and
their $1\sigma$ uncertainties are listed in Table~3 and the resulting
PHOEBE model is shown in Fig.~3. The fundamental parameters of the
components are given in Table~4. We derived masses of
$21.03\pm0.34$~\MS\ and $7.64\pm0.09$~\MS\ and radii of
$9.65\pm0.08$~\RS\ and $9.20\pm0.03$~\RS, for the primary and the
secondary component, respectively. The surface temperature of the
secondary is measured to be $T_{\rm eff_2}=21\,800\pm1200$~K, which is
marginally consistent with the poorly determined spectroscopic one.

\begin{figure}[p]
\vglue-3mm
  \centerline{\includegraphics[width=9.5cm]{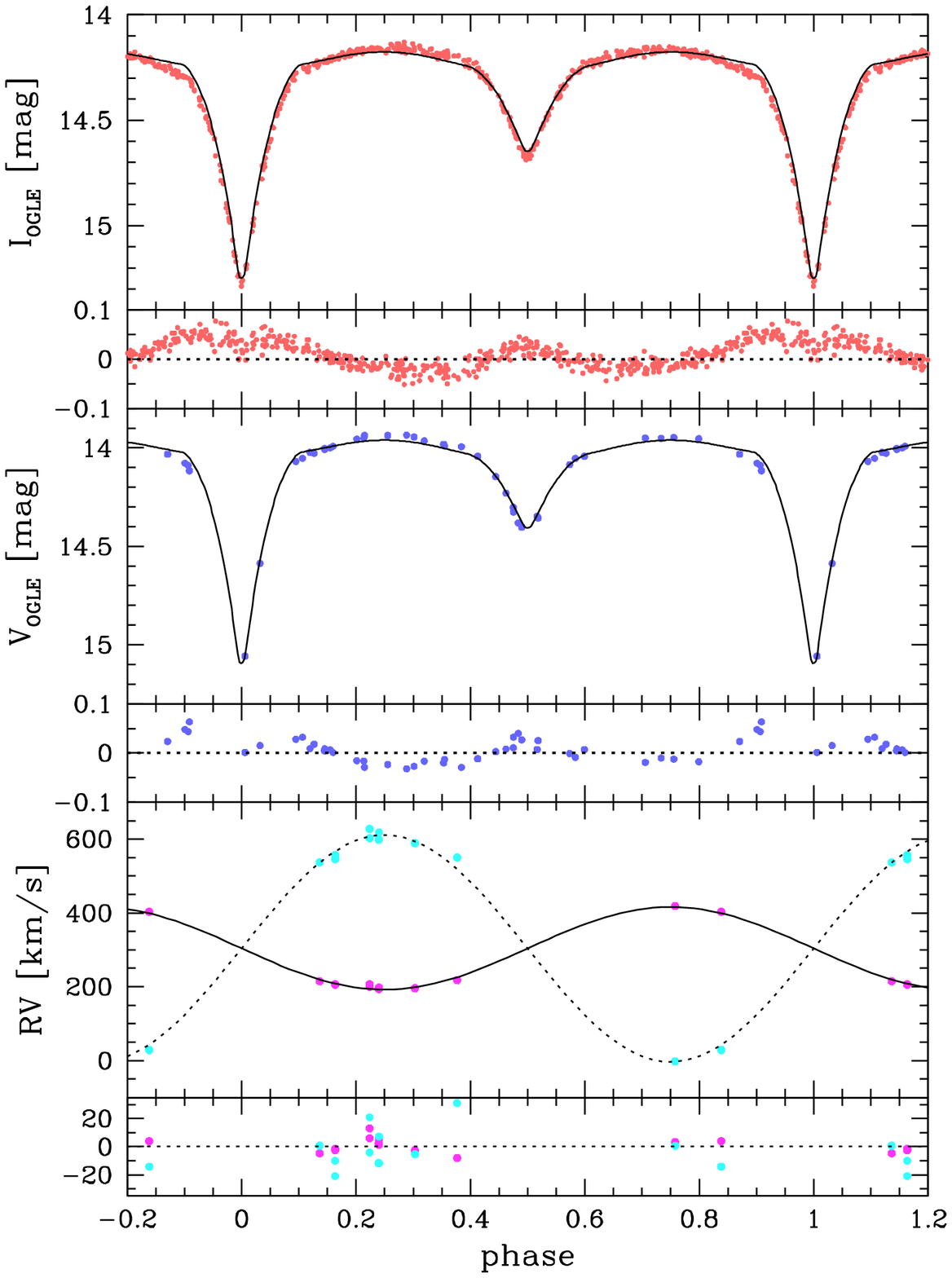}}
  \FigCap{Light curves and radial velocity curves of the best fit
    model with PHOEBE. The {\it top} and {\it middle panels} show the
    {\it I}- and {\it V}-band OGLE-III light curves, respectively,
    together with residuals. The {\it bottom panel} shows the RV curve
    of the primary (magenta) and the secondary (cyan) with the best
    fit model for the primary (solid line) and secondary (dotted
    line), and the residuals.}
\vskip5mm
  \centerline{\includegraphics[width=10cm]{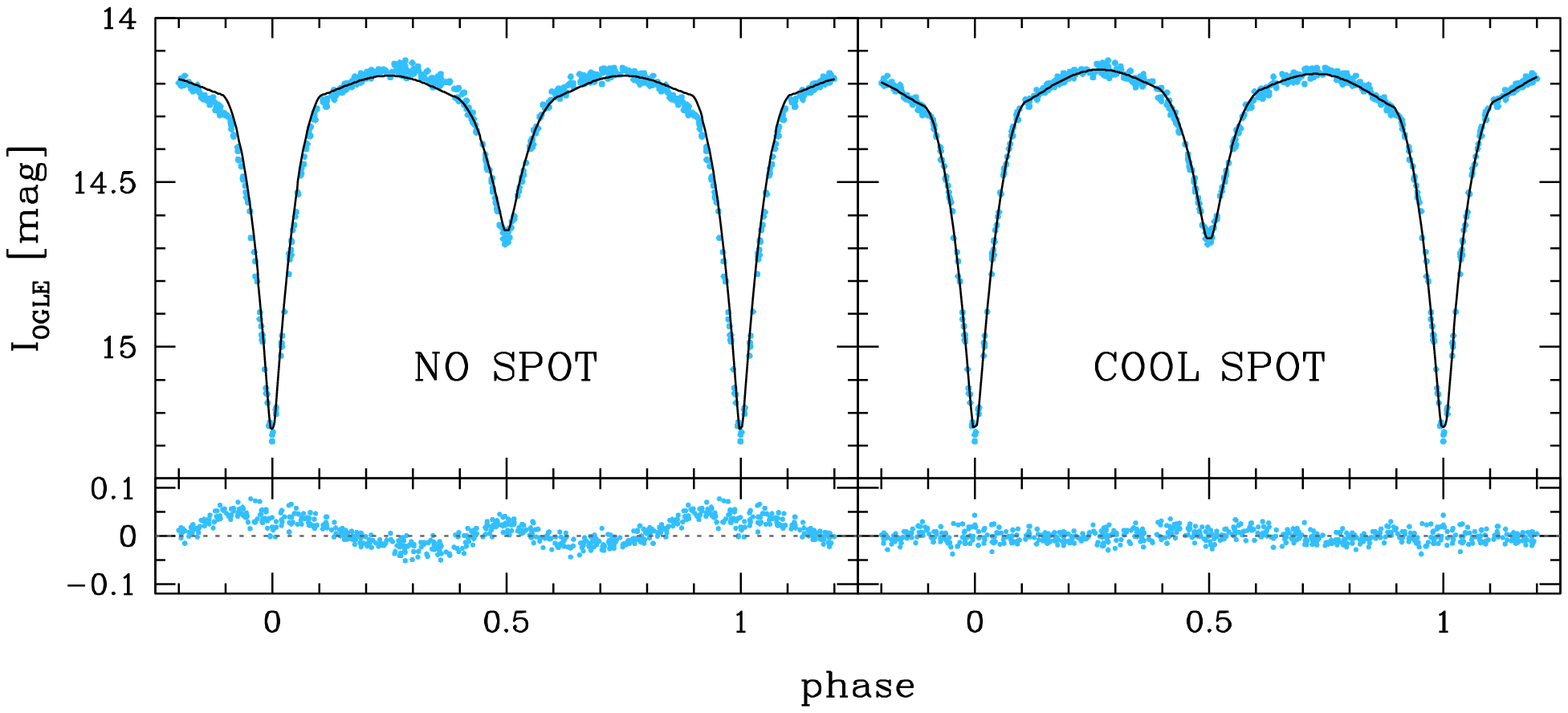}}
  \FigCap{Comparison of two light curve models from PHOEBE -- without
    a cool spot ({\it left panel}) and with the spot ({\it right
      panel}). Blue dots in the top panels show the OGLE-III {\it
      I}-band light curve, while the black line denotes the
    model. {\it Bottom panels} show the residuals between the data and the
    model.}
\end{figure}
\renewcommand{\arraystretch}{1.2}
\MakeTableee{|lr|c|c|}{12.5cm}{Parameters of the OGLE-LMC-ECL-09937 system measured with PHOEBE}
{\hline
Parameter                       & Unit          & Non-Spot Model              & Spot Model$^{(a)}$            \\
\hline
Period $P$                      & [d]           & \multicolumn{2}{|c|}{$3.765229 \pm 0.000003$}         \\
HJD$_0$                         & [d]           & \multicolumn{2}{|c|}{$2453637.48739 \pm 0.0002$} \\
\hline
Semimajor axis, $a$             & [\RS]         &  $31.2 \pm 0.1$       &  $31.2 \pm 0.1$       \\
Systemic velocity, $\gamma$     & [km/s]        &  $ 304 \pm  1$        &  $ 304 \pm  1$        \\
Inclination, $i$                & [deg]         &  $90.0 \pm 0.4$       &  $89.6 \pm 0.4$       \\
Mass ratio, $q$                 &               &  $0.364 \pm 0.004$    &  $0.362 \pm 0.004$    \\
Surface potential of the primary, $\Omega_1$ &  &  $3.64 \pm 0.02 $     &  $3.55 \pm 0.02 $     \\
Flux ratio in the $V-$band      &               &  $2.25 \pm 0.05$      &  $1.90 \pm 0.02$      \\
Flux ratio in the $I-$band      &               &  $2.08 \pm 0.04$      &  $1.80 \pm 0.02$      \\
\hline
\noalign{\vskip5pt}
\multicolumn{4}{p{11.5cm}}{$(a)$ A spot with longitude 190\arcd\!, a radius of 70\arcd\!, and a temperature 0.86 of the stellar surface.}
}
\def\arraystretch{1.0}

\subsection{Spot model}

\vspace*{7pt}
The resulting best-fit light curve model from PHOEBE shows room for
further improvement (see Fig.~3 and the left panel of Fig.~4). The
suppression of the photometric baseline in both the {\it I}- and {\it
V}-band light curves prior to the occultation of the primary implies
the presence of an additional cool component. Since spots are expected
in stars of later types than the one studied here, we attribute the
observed fluctuation in the light curve to the reflection effect, in
which the hotter primary heats the facing surface of the secondary,
increasing its effective temperature. The temperature of the
unreflected region on the secondary (\ie the projected surface of the
secondary at phase zero) would then indicate the actual $T_{\rm
eff_2}$ value.

\begin{figure}[htb]
\centerline{\includegraphics[width=10cm]{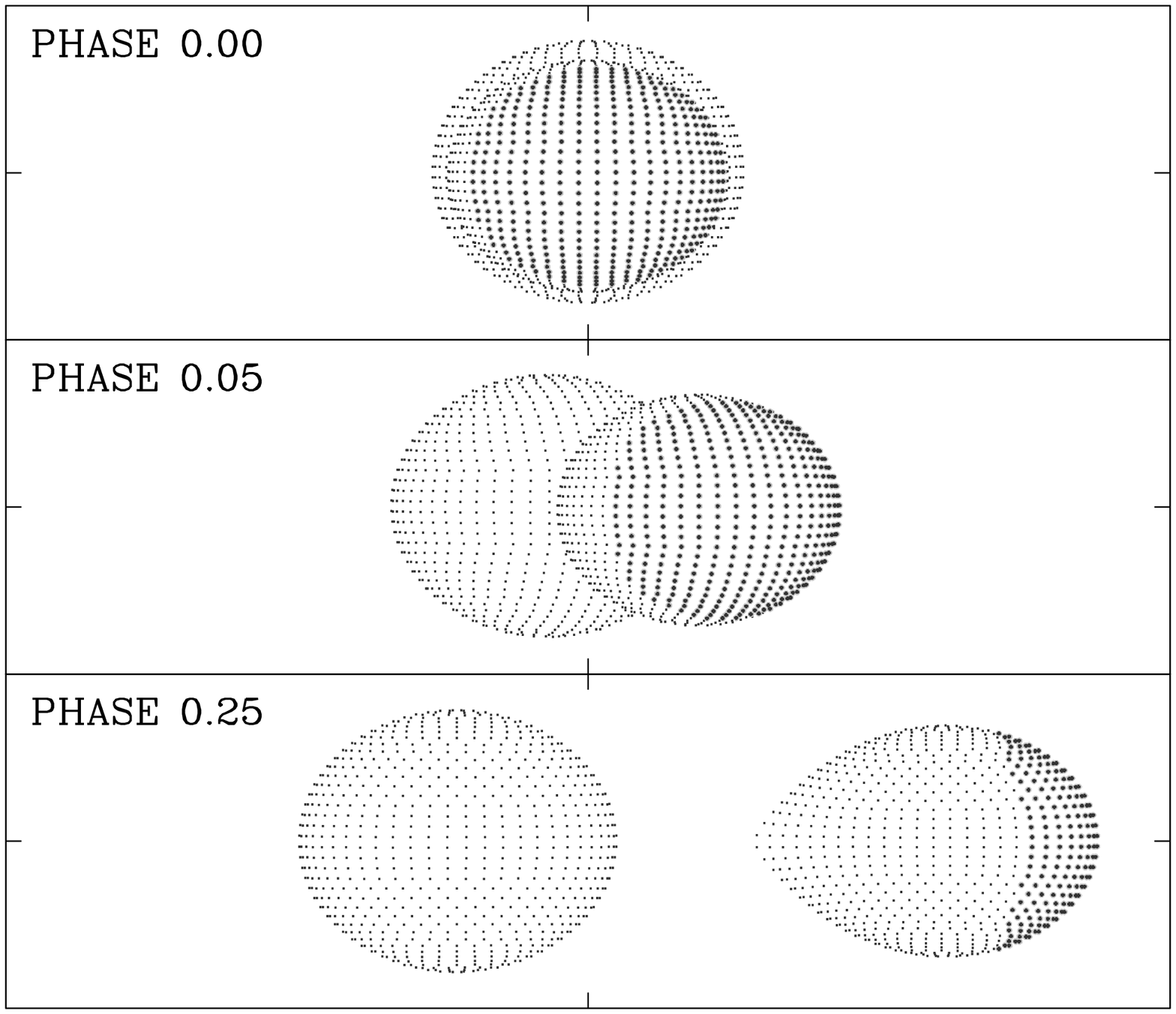}}
\FigCap{Location of the large cool spot on the secondary star is shown
  on a mesh plot imported from PHOEBE at three different orbital
  phases.}
\end{figure}

Once again we used PHOEBE to model the system, this time with the
addition of a large ``cool spot'' on the secondary star, that would
imitate the reflection effect. We assumed an equatorial location of
the spot, \ie the colatitude of 90\arcd\!. The remaining three
parameters of the spot: longitude, radius, and temperature were
determined via multiple fit iterations; each time a different set of
the spot parameters from a wide range of values was chosen, the
observations were fit, and the $\chi^2$ value was returned. The
longitude of the spot of 190\arcd as measured from the L1 point was
found to yield the lowest $\chi^2$ values. The fact that these two
points are not anti-diametrical, is believed to be due to
rotation. For spot radii above 60\arcd\!, the optimization of the fit
models was achieved for a spot temperature of 0.86 of that of the
stellar surface. Assuming a spot radius of 70\arcd\!, spanning the
entire projected surface of the secondary at phase zero (see Fig.~5),
we fixed the three parameters to the above values and repeated the
iterative fit process described in Section~3.2.

The updated parameters of the model are also listed in Table~3 with
the new model shown in the right panel of Fig.~4. The residuals of the
fit are greatly improved when compared to the non-spot model,
indicating a good model selection. The fundamental parameters of the
updated model with the spot are given in the right column of
Table~4. The updated masses are $21.04\pm0.34$~\MS\ and
$7.61\pm0.09$~\MS\ for the primary and the secondary component,
respectively, which are well consistent, within their 1--2\%
precision, with those derived from the previously discussed non-spot
model.

\def\arraystretch{1.2}
\MakeTableee{|lr@{\hspace{30pt}}|@{\hspace{20pt}}c@{\hspace{20pt}}|c|}{12.5cm}{Fundamental parameters of the components of OGLE-LMC-ECL-09937 measured with PHOEBE}
{\hline
Parameter                       & Unit     & Non-Spot Model              & Spot Model$^{(a)}$ \\
\hline
Mass of the primary, $M_1$      & [\MS]         &  $21.03 \pm 0.34 $    &  $21.04 \pm 0.34 $    \\
Mass of the secondary, $M_2$    & [\MS]         &  $7.64 \pm 0.09 $     &  $7.61 \pm 0.09 $     \\
Radius of the primary, $R_1$    & [\RS]         &  $9.65 \pm 0.08$      &  $9.93 \pm 0.06$     \\
Radius of the secondary, $R_2$  & [\RS]         &  $9.20 \pm 0.03$      &  $9.18 \pm 0.04$      \\
$T_{\rm eff_1}$                 & [K]           &  $32000 \pm 1600 $    &  $32000 \pm 1600$     \\
$T_{\rm eff_2}$                 & [K]           &  $21800 \pm 1200 $    &  $21100 \pm 1400$$^{(b)}$     \\
$\log(g_1)$                     &               &  $3.791 \pm 0.008$    &  $3.767 \pm 0.006$    \\
$\log(g_2)$                     &               &  $3.394 \pm 0.002$    &  $3.393 \pm 0.002$    \\
$\log(L_1/\LS)$                 &               &  $ 4.94 \pm 0.09$       &  $ 4.96 \pm 0.09$      \\
$\log(L_2/\LS)$                 &               &  $ 4.23 \pm 0.10$       &  $ 4.17 \pm 0.10$      \\
\hline
\noalign{\vskip5pt}
\multicolumn{4}{p{12.5cm}}{$(a)$ A spot with longitude 190\arcd\!, a radius of 70\arcd\!, and a temperature 0.86 of the stellar surface.}\\
\multicolumn{4}{p{9cm}}{$(b)$ The temperature of the cool spot, \ie unreflected region.}\\
}

\section{Discussion}

By comparing dynamical masses to the predicted evolutionary ones,
eclipsing binaries with spectroscopically measured parameters serve as
a unique tool for evaluating the current theoretical models. We
employed stellar evolutionary models and isochrones from MESA
Isochrones and Stellar Tracks (MIST) v.1.1. (Dotter 2016, Choi \etal
2016, Paxton \etal 2011), assuming rotation at the 40\% of the
critical velocity. The metallicity of the models was set to the
average value of ${\rm [Fe/H]}=-0.39$~dex for the LMC, which is
measured from the calibration of the spectroscopic data of Red Giant
Branch stars (Choudhury \etal 2016).  Stellar tracks for masses of
8~\MS to 24~\MS\ and isochrones of 1~Myr, 3~Myr, 5~Myr, 10~Myr, and
30~Myr, are shown on the Hertzsprung-Russell diagram of Fig.~6. The
black squares show the location of the components on the diagram, as
inferred from our modeling with PHOEBE with the addition of the cool,
unreflected region.

\begin{figure}[htb]
  \centerline{\includegraphics[width=11cm, bb=0 133 570 630, clip=]{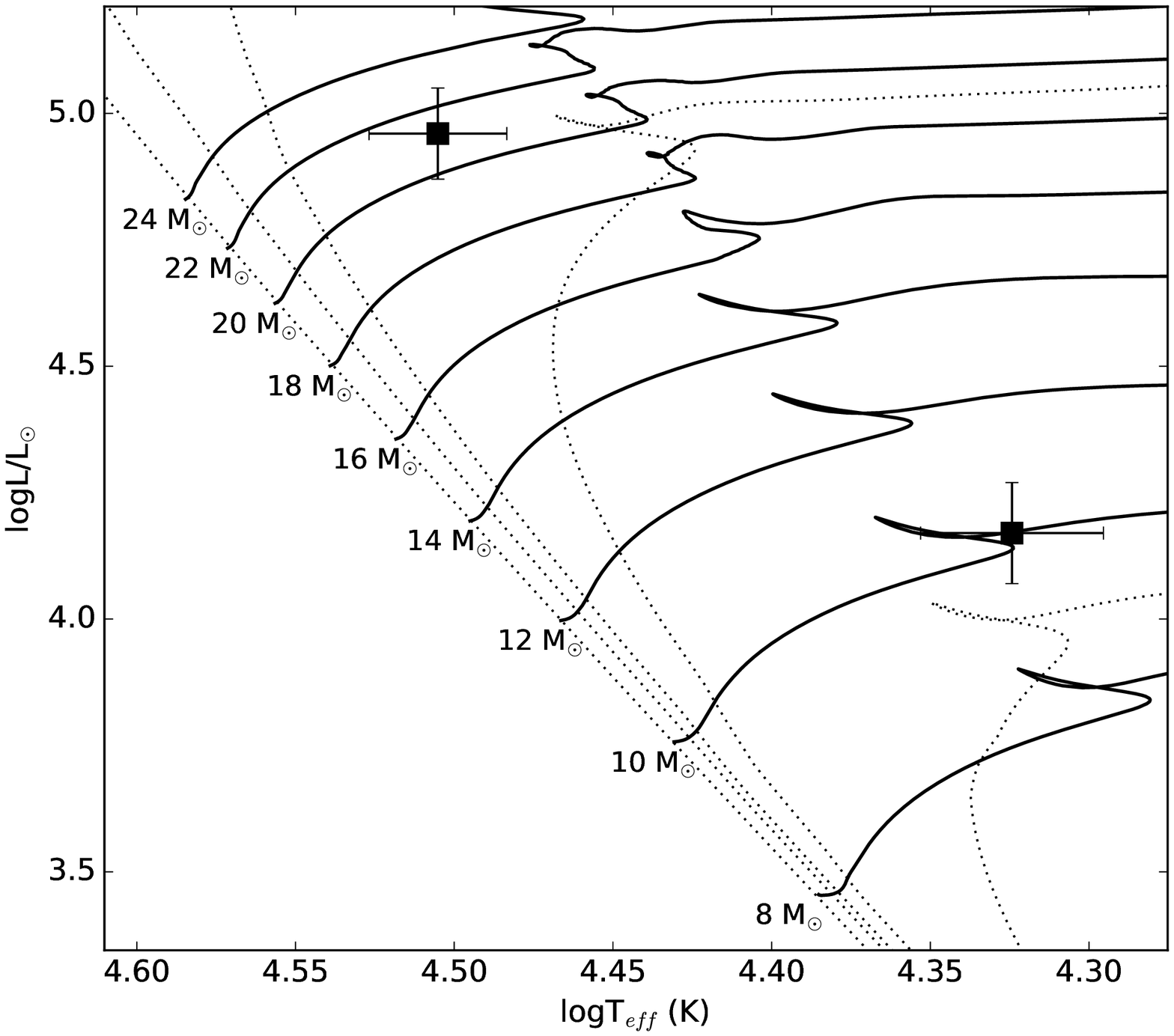}}
  \FigCap{Hertzsprung--Russell diagram for the components of
    OGLE-LMC-ECL-09937 (black squares) studied with PHOEBE. We show MESA
    evolutionary stellar tracks for initial masses 8~\MS\ to 24~\MS,
    assuming rotation at 40\% of the critical velocity. Isochrones of
    1~Myr, 3~Myr, 5~Myr, 10~Myr, and 30~Myr are shown with dotted
    lines.}
\end{figure}

We find a good agreement between the evolutionary and dynamical mass
of 21~\MS\ for the primary component, which appears to reside well
within the main-sequence strip. The secondary component is found to be
$\approx30\%$ overluminous for its mass and further indicates an evolved
state not far beyond the main sequence. As the cooler, fainter, and
less massive secondary is found to fill the Roche lobe, it is
reasonable to consider the system as an Algol-type binary (Giuricin
\etal 1983). Accordingly, the secondary should have been initially a
massive early-type star that underwent a rapid mass transfer to its
companion, which is now the more massive and luminous primary (Peters
2001). The agreement between the primary mass and its theoretical
single-evolved counterpart could imply that the gainer had enough time
to stabilize to its present state and that the mass transfer is
currently absent.

In an attempt to detect possible changes in the orbital period, that
would suggest an ongoing mass transfer, we calculated the orbital
period of the system separately for each OGLE observing season. In
order to extend the baseline of the observations, we supplemented the
OGLE-III {\it I}-band light curve with the OGLE-IV (Udalski \etal
2015) data from the catalog of eclipsing binaries by Pawlak \etal
(2016). The combined dataset spans over 14 years of observations
(14~observing seasons). Furthermore, we downloaded the {\it V}-band
light curve of OGLE-LMC-ECL-09937 from the All-Sky Automated Survey
for Supernovae (ASAS-SN, Shappee \etal 2014) using the ASAS-SN Light
Curve Server (Kochanek \etal 2017), spanning over $3.5$~years of
observations in four observing seasons. This further extended the
baseline to almost 17 years. We calculated an accurate period value
for each of the observing seasons within OGLE-III and ASAS-SN data and
for each two seasons in OGLE-IV data (due to fewer points) with the
{\sc Tatry} code and found no evident variability in the period within
uncertainties.

OGLE-LMC-ECL-09937 is one of the few massive Algol-type systems with
an accurate mass measurement of 2\%. In Fig.~7, we show the
mass--radius plot for the components of the Algol-type systems from
the catalog of Budding \etal (2004) updated with the most recent mass
values for the components of the massive system ET~Tau (Williamon
\etal 2016). Light gray dots represent stars for which there is no
error estimate for the mass, dark gray dots stand for stars for which
the error on the mass is more than 2\%, and large black circles mark the
stars for which masses were determined with up to 2\% accuracy. The
most massive binary in the catalog is RY~Sct, which has had a number
of mass measurements over the past decades, ranging from
$8~\MS+26~\MS$ (Skulskii 1992) to $36~\MS+46~\MS$ (Cowley and
Hutchings 1976). The more recent estimates oscillate around
$\approx10~\MS+30~\MS$ but rarely provide any radii measurement. Hence
we adopt the newest result by Grundstrom \etal (2007) of
$7.1\pm1.2$~\MS\ and $30.0\pm2.1$~\MS. Since the radius is not well
defined, we follow Grundstrom \etal (2007) and adopt the radii of
18~\RS\ and 9.6~\RS, respectively.

Fig.~7 shows that the primary component of OGLE-LMC-ECL-09937 (black
square) is not only the second most massive known star in an Algol-type
system, but it is also the most massive Algol component with $\leq2\%$
accurate mass measurements. It is also one of the few massive stars with
accurate physical parameters, making it an important ingredient of the
stellar mass distribution and an excellent target for verifying stellar
evolution scenarios of semi-detached systems.

\begin{figure}[htb]
  \centerline{\includegraphics[width=10cm]{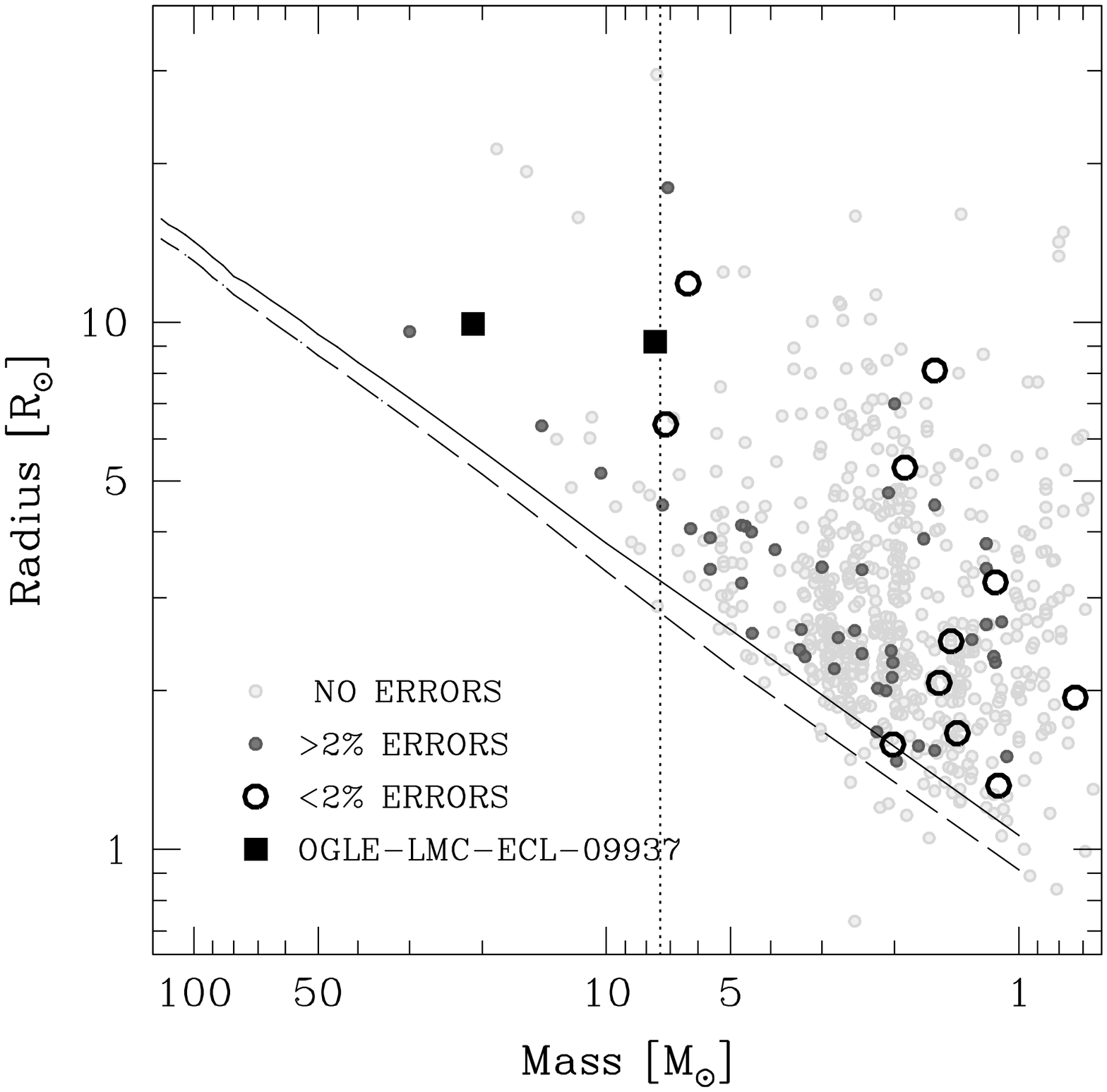}}
  \FigCap{Mass--radius plot for the components of known Algol-type
    systems. Light gray dots represent stars, for which there is no error
    estimate for the mass, dark gray dots represent stars, for which the
    error on the mass is more than 2\%, and large black circles mark the
    stars for which masses were determined with up to 2\% accuracy. The
    black square represents masses of the components of
    OGLE-LMC-ECL-09937. The solid line denotes the zero-age main sequence
    at the Galactic metallicity and the dashed at the LMC metallicity,
    taken from MIST (Dotter 2016, Choi \etal 2016, Paxton \etal 2011). The
    vertical dotted line drawn at $M=7.5$~\MS\ represents the limit above
    which there are no Algol-type component masses known with $\leq2\%$
    uncertainty.}
\end{figure}

\section{Summary}

This paper presents a detailed analysis of the photometric and
spectroscopic data on the semi-detached eclipsing binary system
OGLE-LMC-ECL-09937. The spectra of the system show lines of both
components, allowing to determine both radial velocity curves, which
together with the excellent quality OGLE {\it I}- and {\it V}-band
light curves, allowed us to obtain an accurate model of the system
using PHOEBE. The system is composed of a hot, massive and luminous
primary star of an early B type, and a more evolved, but less massive
and luminous secondary, implying an Algol-type system that underwent a
mass transfer episode.

Further modeling of the binary shows that the light curve fit is
greatly improved when adding a large cool region on the secondary,
which is attributed to the reflection effect. The physical parameters
of the system derived from the non-spotted and cool spot models are
consistent within their errors. We derive masses of
$21.04\pm0.34$~\MS\ and $7.61\pm0.09$~\MS\ and radii of
$9.93\pm0.06$~\RS\ and $9.18\pm0.04$~\RS, for the primary and the
secondary component, respectively.

The $<2\%$ accuracy of the mass and radius measurements make
OGLE-LMC-ECL-09937 an important contribution to the sparsely populated
high-mass end of the stellar mass distribution, and an interesting
object for stellar evolution studies, as a potential progenitor of a
binary system composed of two neutron stars.


\Acknow{
We thank Kris Stanek for initiating and supporting this project.
D.M.S and M.K. are supported by the Polish Ministry of Science and Higher 
Education under the grant ``Iuventus Plus'' No. 0420/IP3/2015/73. D.M.S. is 
supported by the Polish National Science Center (NCN) under the grant no. 
2013/11/D/ST9/03445 and the NCN grant MAESTRO 2014/14/A/ST9/00121 to
Andrzej Udalski. Support for J.L.P. is provided in part by FONDECYT through
the grant 1151445 and by the Ministry of Economy, Development, and Tourism's
Millennium Science Initiative through grant IC120009, awarded to The Millennium
Institute of Astrophysics, MAS.}



\begin{references}
\refitem{Bernstein, R., Shectman, S. A., Gunnels, S. M., Mochnacki, S., Athey, A. E.}{2003}{SPIE}{4841}{1694}
\refitem{Bonanos, A. Z., \etal}{2006}{\ApJ}{652}{313}
\refitem{Bonanos, A. Z.}{2009}{\ApJ}{691}{407}
\refitem{Bonanos, A. Z., \etal}{2009}{\AJ}{138}{1003}
\refitem{Budding, E., Erdem, A., \c{C}i\c{c}ek, C., Bulut, I., Soydugan, F., Soydugan, E., Baki\c{s}, V., Demircan, O.}{2004}{\AA}{417}{263}
\refitem{Choi, J., Dotter, A., Conroy, C., Cantiello, M., Paxton, B., Johnson, B.}{2016}{\ApJ}{823}{102}
\refitem{Choudhury, S., Subramaniam, A., Cole, A.}{2016}{\MNRAS}{455}{1855}
\refitem{Conti, P.}{1971}{\ApJ}{170}{325}
\refitem{Cowley, A. P., Hutchings, J. B.}{1976}{\PASP}{88}{456}
\refitem{Dotter, A.}{2016}{\ApJ}{222}{8}
\refitem{Giuricin, G., Mardirossian, F., Mezzetti, M.}{1983}{ApJS}{52}{35}
\refitem{Graczyk, D., \etal}{2011}{\Acta}{61}{103}
\refitem{Grundstrom, E. D., Gies, D. R., Hillwig, T. C., McSwain, M. V., Smith, N., Gehrz, R. D., Stahl, O., Kaufer, A.}{2007}{ApJ}{667}{505}
\refitem{Hubeny, I., Lanz, T.}{1995}{\ApJ}{439}{875}
\refitem{Kochanek, C. S. \etal}{2017}{\PASP}{129}{4502}
\refitem{Massey, P., Lang, C. C., Degioia-Eastwood, K., Garmany, C. D.}{1995}{\ApJ}{438}{188}
\refitem{Pawlak, M., \etal}{2016}{\Acta}{66}{421}
\refitem{Paxton, B., Bildsten, L., Dotter, A., Herwig, F., Lesaffre, P., Timmes, F.}{2011}{\ApJ}{192}{3}
\refitem{Peters, G. J.}{2001}{ASSL}{264}{79}
\refitem{Pojma\'nski, G.}{1997}{\Acta}{47}{467}
\refitem{Pojma\'nski, G.}{2002}{\Acta}{52}{397}
\refitem{Pr\v{s}a, A., and Zwitter, T.}{2005}{\ApJ}{628}{426}
\refitem{Puls, J., Urbaneja, M. A., Venero, R., Repolust, T., Springmann, U., Jokuthy, A., Mokiem, M. R.}{2005}{\AA}{435}{669}
\refitem{Pych, W.}{2004}{\PASP}{116}{148}
\refitem{Sana, H. \etal}{2012}{Sci}{337}{444}
\refitem{Santolaya-Rey, A. E., Puls, J., Herrero, A.}{1997}{\AA}{323}{488}
\refitem{Schwarzenberg-Czerny, A.}{1996}{\ApJ}{460}{L107}
\refitem{Shappee, B. J. \etal}{2014}{\ApJ}{788}{48}
\refitem{Skulskii, M. Y.}{1992}{SvA}{36}{411}
\refitem{Southworth, J.}{2015}{ASPC}{496}{164}
\refitem{Szczygie\l, D. M., Stanek, K. Z., Bonanos, A. Z., Pojma\'nski, G., Pilecki, B., Prieto, J. L.}{2010}{\AJ}{140}{14}
\refitem{Tody, D.}{1986}{SPIE}{627}{733}
\refitem{Tody, D.}{1993}{ASPC}{52}{173}
\refitem{Toma, K., Yoon, S-C., Bromm, V.}{2016}{SSRv}{202}{159}
\refitem{Torres, G., Andersen, J., Gim\'enez, A.}{2010}{AARv}{18}{67}
\refitem{Udalski, A.}{2003}{\Acta}{53}{291}
\refitem{Udalski, A., Szyma\'nski, M. K., Soszy\'nski, I., Poleski, R.}{2008}{\Acta}{58}{69}
\refitem{Udalski, A., Szyma\'nski, M. K., and Szyma\'nski, G.}{2015}{\Acta}{65}{1}
\refitem{van Hamme, W.}{1993}{\AJ}{106}{2096}
\refitem{Walborn, N. R., Fitzpatrick, E. L.}{1990}{PASP}{102}{379}
\refitem{Williamon, R. \etal}{2016}{\PASP}{128}{4202}
\refitem{Wilson, R. E., Devinney, E. J.}{1971}{\ApJ}{166}{605}
\refitem{van Hamme, W.}{1993}{AJ}{106}{2096}

\end{references}
\end{document}